# Questioning van der Waals Epitaxy of Non-Layered Materials on Mica: The Case of ScN

Susmita Chowdhury,[1,*] Faezeh Alijan Farzad Lahiji,[1,2,3] Mikael Ottoson,[1] Olivier Donzel-Gargand,[4] Robert J. W. Frost,[5] Martin Magnuson,[2] Ganpati Ramanath,[1,6,7] Arnaud le Febvrier,[1] and Per Eklund[1,2,7,*]

[1]*Inorganic Chemistry, Department of Chemistry - Ångström Laboratory, Uppsala University, Box 538, SE-751 21 Uppsala, Sweden*

[2]*Thin Film Physics Division, Department of Physics, Chemistry and Biology (IFM), Linköping University, SE-581 83 Linköping, Sweden*

[3]*Materials Chemistry, RWTH Aachen University, Kope. 10, D-52074, Aachen, Germany*

[4]*Division of Solar Cell Technology, Department of Materials Science and Engineering, Uppsala University, SE-751 21 Uppsala, Sweden*

[5]*Division of Materials Physics, Department of Physics and Astronomy, Uppsala University, SE-751 21 Uppsala, Sweden*

[6]*Department of Materials Science and Engineering, Rensselaer Polytechnic Institute, Troy, NY 12180, USA*

[7]*Wallenberg Initiative Materials Science for Sustainability, Department of Chemistry, Uppsala University, 751 21 Uppsala, Sweden*

[*] Corresponding authors: susmita.chowdhury@kemi.uu.se, per.eklund@kemi.uu.se

**Abstract**

Growing stress-free epitaxial films by van der Waals epitaxy (vdWE) is of interest for realizing flexible optoelectronics and energy devices from freestanding epilayers. However, vdWE of non-layered materials is often presumed or claimed on layered substrates such as mica with inadequate experimental evidence. Here, we demonstrate that the growth of single-domain rocksalt ScN(111) films by sputter deposition on fluorophlogopite mica(001) occurs by conventional epitaxy. X-ray diffraction and electron microscopy reveal the film/substrate epitaxial relationship to be $[\bar{1}01](111)_{ScN}\|[010](001)_{mica}$. Our results indicating strain buildup seen from the dependence of (111) interplanar spacings, and strain relaxation by dislocation generation, question prior claims of vdWE of non-layered metal nitrides on mica. Our findings show that conventional epitaxy should be the default assumption for non-layered materials unless conditions for vdWE are explicitly established.

There is great interest in van der Waals epitaxy (vdWE) of organic and inorganic films on substrates of layered materials held together by vdW bonding, e.g., graphene, mica, hexagonal boron nitride, and molybdenum disulfide[1–3]. Films with weak film-substrate bonding[2,4] and near-zero strain are key features of vdWE besides strong in-plane and out-of-plane textures. vdWE offers the possibility to realize thick epilayers without dislocation generation, unlike conventional epitaxy which involves strong interface bonds that lead to stress buildup with increasing film thickness, and interface dislocations above a critical epilayer thickness. Also, epilayers grown by vdWE can be easily released from the substrate for transfer to other substrates or used as free-standing films.[5] Besides in-plane and out-of-plane texturing of the epilayer, distinguishing features of vdWE are the fact that the film strain does not depend on film thickness and exhibits weak film-substrate interface bonding[6–8]. Claims of vdWE without establishing these necessary conditions have raised concerns about their validity.

Numerous studies have definitively demonstrated vdWE for 2D-layered materials on substrates of 2D-layered materials[5,9,10] such as mica, where layering connotes strong in-plane $sp^2$ bonding and weak out-of-plane vdW bonding. Examples include $Bi_2X_3$ (X = Te and Se)[11], hexagonal Te[12], $MoS_2$[13], orthorhombic α-$MoO_3$[14] and monoclinic $VO_2$[4]. The question is far more complex for non-layered materials (*i.e.*, 3D materials with strong non-directional bonding) on mica, e.g., NiO[15], TiN[16], perovskite $SrTiO_3$[17], wurtzite InN[18], and ZnO[2]. In this context, vdWE has been claimed in $Al_2O_3$[10], monoclinic $MoO_2$[19] on muscovite mica, rocksalt ScN[20] and wurtzite GaN[21] films grown on f-mica. For non-layered transition metal nitrides, the $sp^3d^2$ electronic structure leads to exciting optoelectronic and magnetic properties[3,16,22] underpinned by strong covalent, ionic and/or metallic bonds[23,24] that favor conventional epitaxy. Despite this, as noted above, vdWE has been assumed or claimed on layered substrates featuring vdW bonds. Thus, the question still remains open to what extent vdWE can be achieved for NaCl rocksalt type transition metal nitrides with layered materials serving as a template.

Here, we demonstrate that rocksalt ScN grows on fluorophlogopite mica (f-mica, c2/m space group) by conventional epitaxy. ScN has attractive thermoelectric[25], piezoelectric[26] and plasmonic properties[27] for harvesting electricity from waste heat and motion. Monoclinic f-mica is mechanically flexible, optically transparent in the UV to IR range, thermally stable between -100°C to 1100°C and therefore a flame retardant, and resilient against chemical attack from concentrated acids, such as, HCl and $H_2SO_4$. This variant of mica specified by $KMg_3(AlSi_3O_{10})F_2$ has a higher purity and thermomechanical stability[28] than muscovite mica[29] and features ionic and covalent bonding within layers of $(SiAl)O_4$ tetrahedra and $Mg(OF)_6$ octahedra, and vdW bonding between the *00l* basal plane of K atoms and the $(SiAl)O_4/Mg(OF)_6/(SiAl)O_4$ stack[30].

ToF-ERDA results indicate stoichiometric ScN film seen from an elemental ratio of N/Sc =0.95±0.01. Adventitious impurities include ~4.5 at.% oxygen in the film due to low Sc-O bond formation enthalpy[31], ~1.1 at. % carbon and ~0.2 at. % hydrogen in the bulk of the film (SI Section S3)[32].

X-ray diffractograms showing out-of-plane 111 and 222 reflections from ScN, besides f-mica(00*l*) peaks (Fig. 1a), indicate ScN(111) planes stacked along [111]. High-resolution scans (Fig. 1b) reveal a monotonic decrease in (111) and (222) interplanar spacings $d_{111}$ and $d_{222}$ with increasing film thickness $t_{film}$ (Fig. 1c). The $d_{222}$ value for the thinnest ScN film ($t_{film}$ = 8 nm) could not be accurately ascertained due to the large peak width. Thicker ScN films, *i.e.*, $14 \leq t_{film} \leq 138$ nm) exhibit 2.604 (±0.004) $\leq d_{111} \leq$ 2.632 (±0.006) Å range and 1.302 (±0.004) $\leq d_{222} \leq$ 1.305 (±0.005) Å.

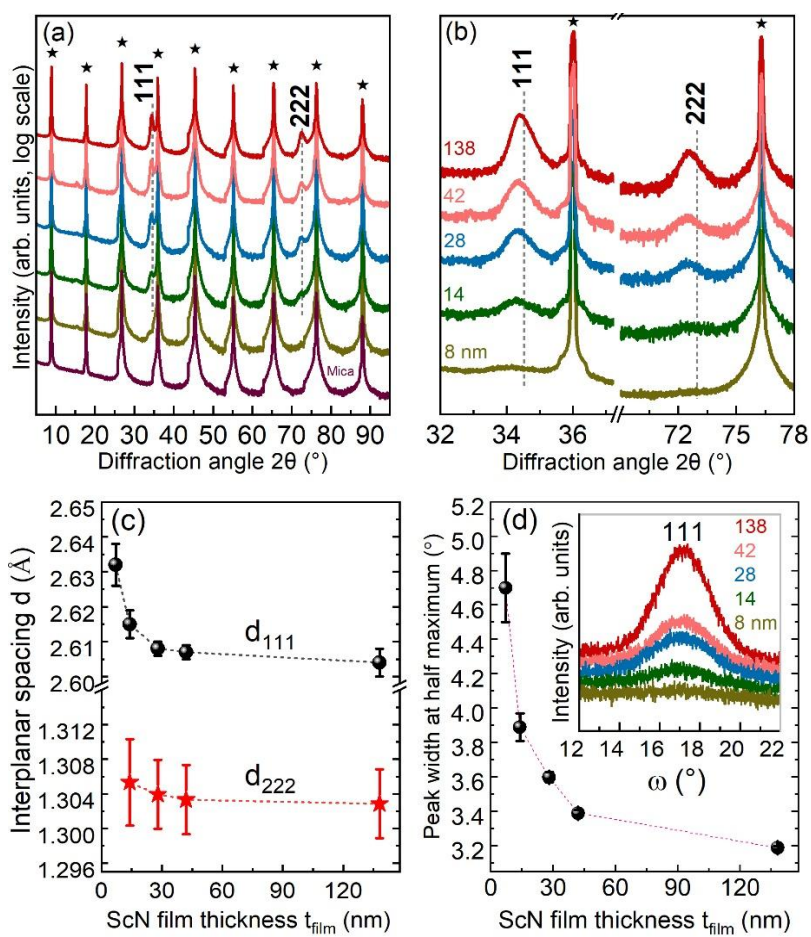

Fig. 1. Representative X-ray diffractograms (a) with a Cu Kα beam and (b) high-resolution scans obtained from ScN films of different film thicknesses on f-mica. (c) Interplanar spacings

of 111 and 222, $d_{111}$ and $d_{222}$ plotted versus ScN film thickness $t_{film}$. (d) Full-width-at-half-maximum (FWHM) of the (111) peak versus $t_{film}$ using data from ω scan rocking curves (inset).

The ScN(111) peak width determined from rocking curve ω-scans decreases with increasing $t_{film}$ (Fig. 1d). The full-width half-maximum (FWHM) peak width of ~3.19° even for the thickest ScN film ($t_{film}$ = 138 nm) could be from misoriented crystal domains and high defect density. The gradual decrease in peak width from 4.70° (±0.20) to 3.39° (±0.02) with increasing $t_{film}$, suggests domain coalescence and defect annihilation.

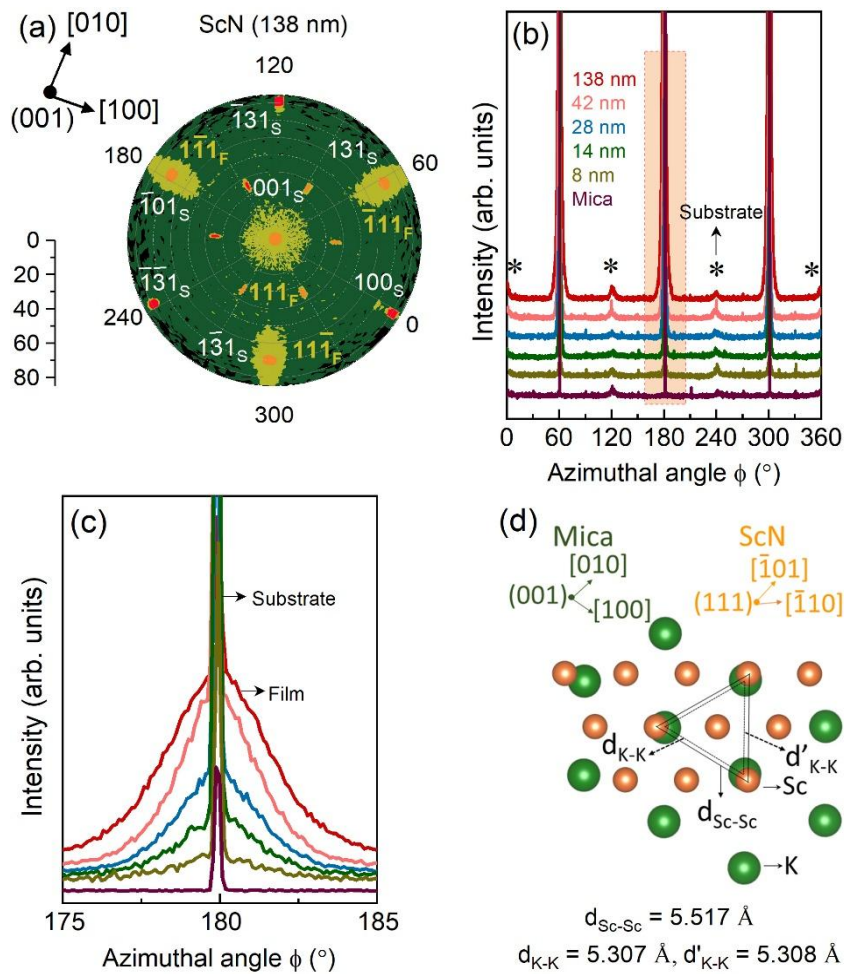

Fig. 2. (a) Pole figures from ScN(111) reflection for 2θ = 34.35° from the thickest ScN film. (b) ϕ scans showing the ScN(111) pole peaks and the f-mica(001) peaks. (c) A magnified view of ϕ scans around ϕ ≈ 180° depicts the high width of ScN(111) poles films in contrary to the narrow pole from f-mica(001). (d) Schematics depicting atomic registry between epitaxial ScN(111) and f-mica (001) with green spheres depicting K in mica and red spheres showing Sc in ScN.

Pole figures from the thickest ScN(111) film (Fig. 2a) for Bragg angle 2θ = 34.35° show two sets of poles: the sharp poles (suffixed s) from the substrate and relatively broad features for the film (suffixed f), *e.g.*, at ψ = 0° and at ~ 70°. The pole cluster around ψ = ϕ = 0° connotes ScN(111). The poles at ψ ~ 70° and ϕ ≈ 60°, 180° and 300° correspond to ScN($\bar{1}$11), (1$\bar{1}$1) and (11$\bar{1}$). All the ScN poles are broad, indicating high crystal mosaicity and defects. The six other poles at ψ ~ 34.7° separated by Δϕ ≈ 60°, and the three poles each at ψ ~ 71° and at ψ ~ 79° with Δϕ ≈ 120° arise from f-mica (SI Fig. S2)[32]. These f-mica poles from intense 131, $\bar{1}$01 and 1$\bar{3}$1, and faint $\bar{1}$31, $\bar{1}\bar{3}$1 and 100 reflections are all narrow. Azimuthal ϕ scans at ψ ~ 70° distinguish the broad poles from ScN and the narrow poles from f-mica (Fig. 2b). A magnified view for ϕ ~ 180° (Fig. 2c) confirms broad ScN poles characterized by Δϕ ~ ±5° for different film thicknesses while the f-mica poles are about tenfold narrower.

The threefold symmetry of ScN{111} poles indicate that ScN(111) is related epitaxially to f-mica(001) by [$\bar{1}$01](111)$_{ScN}$||[010](001)$_{mica}$. This film/substrate atomic registry is captured schematically (Fig. 2d). For simplicity, only K atoms from mica(001) planes[30] and Sc atoms from ScN(111) planes are represented. The f-mica structure informed by PCPDF #04-009-4162 (f-mica: a = 5.307 Å, b = 9.195 Å, c = 10.134 Å, β = 100.08°) and that of ScN by PCPDF#04-001-1145 (a = 4.505 Å), indicate an in-plane lattice mismatch of ($d_{K-K}-d_{Sc-Sc}$)/$d_{K-K}$ ≈ -3.95% and ($d'_{K-K}-d_{Sc-Sc}$)/$d'_{K-K}$ ≈ -3.78% along the diagonal.

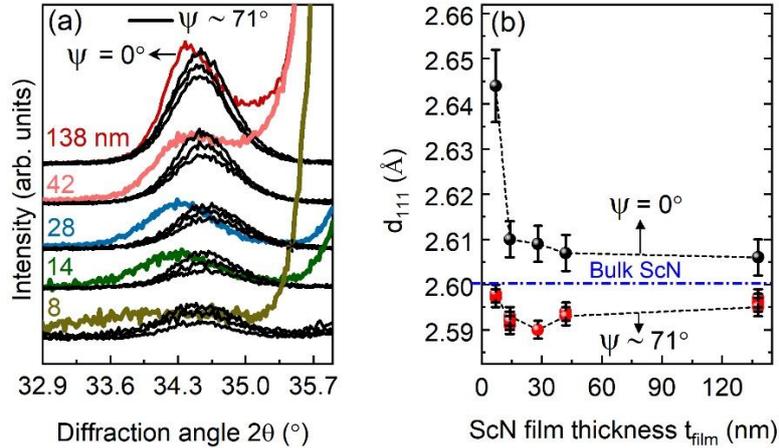

Fig. 3. (a) ScN{111} peaks from high resolution scans around ψ = 0° and three additional locations with ψ ~ 71° plotted as a function of the diffraction angle 2θ. (b) ScN(111) interplanar-spacings for ψ = 0° and ψ ~ 71° plotted versus ScN film thickness $t_{film}$. The horizontal line depicting the bulk ScN(111) interplanar spacing $d_{111}$ = 2.60 Å from PCPDF #04-001-1145.

The above results indicate that sputter-deposited ScN(111) epilayers on f-mica (00*l*) feature a single domain of (111) planes with ABCABC stacking (confirmed by electron diffraction, SI section S4), wherein the lattice mismatch drives mosaicity and defects. This result is contrary to report of twin (111) domains with ABCABC and ACBACB stacking reported for molecular beam epitaxy (MBE)-grown ScN on f-mica[20]. This difference could be from vastly different energetics of surface adatom interactions that could alter crystal growth evolution paths[33]. Close examination of the Bragg angles reveals that the three non-central $\{\bar{1}11\}$ planes ($\psi \sim 71°$) are similar within experimental uncertainty, but larger than that of the central (111) peak at $\psi = 0°$ (Fig. 3a). This difference indicates in-plane biaxial compressive film strain. The decreasing difference with increasing film thickness $t_{film}$ (Fig. 3b) implies strain relaxation.

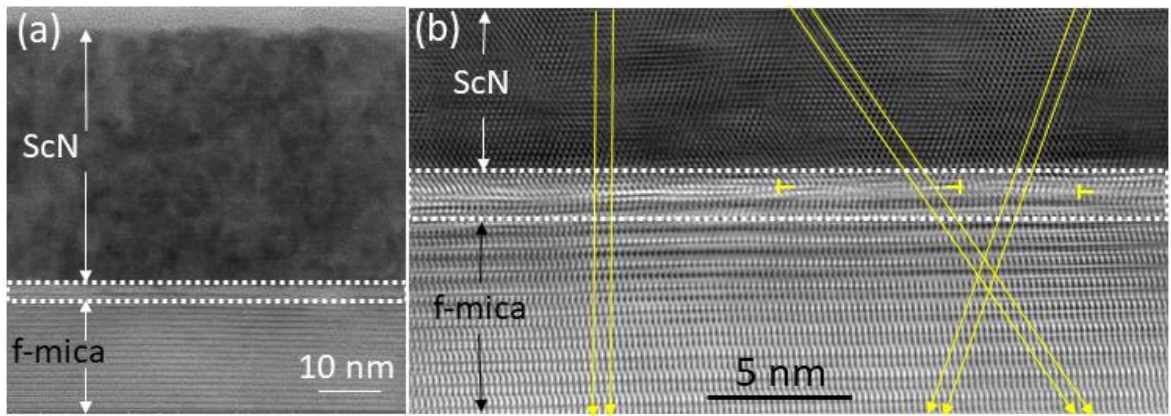

Fig. 4. (a) Drift corrected frame integrated bright field STEM micrograph from a ScN(111) grown on f-mica(001). (b) A magnified view showing the atomistic details of the distorted mica up to a few atomic planes, with ⊢ and ⊣ indicating extra half-planes from edge dislocations, while diagonal and vertical arrows show atom on atom registry *i.e.*, epitaxy.

Drift corrected frame integrated bright-field STEM images (Figs. 4a-b) confirm ScN(111) epitaxy on mica(00*l*) and indicate a sharp interface of ScN (dark contrast over 1-2 atomic planes only) on mica (white contrast). Atomic registry across the ScN film and f-mica is illustrated by diagonal and vertical arrows. The distorted lattice contrast at the mica surface up to a few atomic planes (~2 nm) arises from edge dislocations indicated by ⊢ and ⊣ connoting extra half-planes and crystal plane rotation away from the electron-beam axis. This is caused by the induced strain. The dislocation formation to relax film strain is characteristic of conventional epitaxy, and refutes vdWE.

Strain buildup in tens-of-nanometers-thick ScN films and relaxation by dislocation formation are classic features of conventional epitaxy of ScN on f-mica. Our results contradict claims of vdWE of rocksalt ScN[20], wurtzite GaN[21] on f-mica and monoclinic $MoO_2$[19] on muscovite mica because vdWE features weak (1~100 meV atom$^{-1}$) film/substrate bonding that cannot support strain buildup. Moreover, the resilience of MBE ScN epilayers to delamination during cyclic loading[20] suggests conventional epitaxy supported by strong covalent and ionic bonds (1~10 eV atom$^{-1}$)[34] at the film/substrate interface bonding. Further, any claims of free-standing non-layered films detached from mica require evidence by compositional analysis to rule out the presence of remains of the substrate. Unequivocally resolving this issue requires the examination of thickness-dependence of strain evolution to verify or rule out vdWE. Thus, we conclude that monitoring stress evolution should be a necessary condition for any claims of vdWE. In light of this, it is clear that conventional epitaxy should be the default assumption for non-layered materials unless there is compelling evidence of independence of interplanar spacing on film thickness to support vdWE.

In summary, our results show that sputter-deposited ScN thin films on f-mica exhibit conventional epitaxy rather than vdWE. In particular, we find ScN crystals grow in a single-domain epitaxy specified by $[\bar{1}01](111)_{ScN}\|[010](001)_{mica}$. The presence of significant compressive strain in the ScN films, and its dependence on thickness refutes vdWE. These results lead to the conclusion that conventional epitaxy should be the default assumption for non-layered materials on mica, especially in the absence of compelling evidence of strain independence on epilayer thickness to indicate vdWE. These results should be applicable to other rocksalt transition metal nitrides and oxides with non-layered crystal structures.

**Methods**

ScN thin films with thicknesses between ~8 to ~138 nm was grown by direct-current magnetron sputter deposition on f-mica (*00l*) surfaces [Continental Trade, Warszawa, Poland] freshly exfoliated by scotch tape. The substrate temperature was maintained at ~800°C. The substrate holder was rotated at 15 rpm for homogeneous film deposition. A 2-inch 99.9% pure Sc target was sputtered using a 110 W plasma generated by 13 sccm Ar and 31 sccm $N_2$ gas mixture at ≈0.25 Pa. Prior to deposition, the system was degassed to ≤1.3×10$^{-6}$ Pa. The deposition time was adjusted to achieve desired film thicknesses. X-ray reflectivity measurements were carried out to determine the thicknesses of the ScN thin films [section S1 of the supplemental information (SI)][32].

Time-of-flight Elastic Recoil Detection Analysis (ToF-ERDA) measurements were carried out on thickest ScN film using a primary ion beam of $^{127}I^{8+}$, with an energy of 36 MeV, incident at an angle of 22.5° to the sample-surface normal. Recoil ions were detected with a time energy telescope, positioned at an angle of 45° relative to the path of the primary ion beam. Data evaluation was performed using the POTKU software[35] to obtain the depth dependent concentrations of elements present in the sample. Subsequently, the results were further refined through Monte Carlo simulation using MCERD[36]. Details of ToF-ERDA results has been discussed in SI Section S3[32].

Symmetric Bragg-Brentano ω-2θ X-ray diffractograms (XRD) with ω=2θ/2 were acquired using CuK$_\alpha$ radiation powered by 40 mA at 45 kV in an Empyrean, Malvern Panalytical system with multicore optics. The incident beam passed through a 1/4° divergence slit and a Ni filter, and the diffracted beam was detected using a PIXcel 1D detector. The step size Δθ=0.02° and the counting time per step was 197 s. High-resolution diffractograms around the (111) and (222) reflections, and ω scans around the 111 peaks, were obtained using a Malvern Panalytical X'Pert Materials Research diffractometer with CuK$_{\alpha 1}$ source. In-line scanning detection was used for θ-2θ scans, and a receiving slit with a resolution of 0.02° was used for ω scans. Pole figure scans for ScN{111} reflections were acquired for 2θ=34.4° to confirm ScN epitaxy on f-mica [SI, Section S2]. Since superimposed poles from the substrate and the film were difficult to distinguish, we acquired ϕ scans in the Malvern Panalytical X'Pert MRD system with a 4×4 mm$^2$ spot. The point-focused incident beam traversed poly capillary optics and crossed slits. The diffracted beam was collected by a proportional detector. The X-ray diffractograms along {$\bar{1}11$} were also acquired at ψ ≠ 0° for 2θ~34°.

Scanning transmission electron microscopy (STEM) was carried out at 200 kV in a Titan Themis200 on sample cross-section prepared by focused ion beam milling. To protect the sample from possible beam damages, it was first coated with carbon. To improve its conductivity and reduce possible charging during the preparation of the lamella, the whole surface was further coated with AuPd. The lamella was cut and extracted in a Crossbeam550 from Zeiss and polished with Ga$^+$ ions in a Thermofisher Scientific Helios5 instrument. For the final steps, the ion accelerating voltage was decreased gradually from 5 kV to 1 kV to minimize ion-implantation and beam damages. The mica substrate was rapidly degrading during STEM measurement due to its sensitivity to the fast electron beam. It could volatilize within a couple of minutes even when lowering the probe current to 20 pA. Therefore, the measurements presented here were recorded on pristine regions of the lamella, *i.e.*, without prior exposure to the electron beam.

**Acknowledgements**

The authors acknowledge funding from the Swedish Government Strategic Research Area in Materials Science on Functional Materials at Linköping University (Faculty Grant SFO-Mat-LiU No. 2009 00971), the Knut and Alice Wallenberg foundation through the Wallenberg Academy Fellows program (KAW-2020.0196), the Swedish Research Council (VR) under Project No. 2021-03826, 2025-03680 (PE), 2025-03760 (AF), 2025-03705 (MM) and 2024-04996 (GR). MM acknowledges financial support from the Swedish Energy Agency (Grant No. 43606-1) and the Carl Tryggers Foundation (CTS 25:3972, CTS23:2746, CTS20:272). SC thanks Erik Lewin for fruitful discussions and Sanath Kumar Honnali for his help during pole figure measurement. GR acknowledges support from the Wallenberg Initiative Materials Science for Sustainability (WISE) funded by the Knut and Alice Wallenberg Foundation, the US NSF grant CMMI 2135725 through the BRITE program, the Empire State Development's Division of Science, Technology and Innovation Focus Center at RPI (C210117).

Supporting Information

# Questioning van der Waals Epitaxy of Non-Layered Materials on Mica: The Case of ScN

Susmita Chowdhury,[1,*] Faezeh Alijan Farzad Lahiji,[1,2,3] Mikael Ottoson,[1] Olivier Donzel-Gargand,[4] Robert J. W. Frost,[5] Martin Magnuson,[2] Ganpati Ramanath,[1,6,7] Arnaud le Febvrier,[1] and Per Eklund[1,2,7,*]

[1]*Inorganic Chemistry, Department of Chemistry - Ångström Laboratory, Uppsala University, Box 538, SE-751 21 Uppsala, Sweden*

[2]*Thin Film Physics Division, Department of Physics, Chemistry and Biology (IFM), Linköping University, SE-581 83 Linköping, Sweden*

[3]*Materials Chemistry, RWTH Aachen University, Kope. 10, D-52074, Aachen, Germany*

[4]*Division of Solar Cell Technology, Department of Materials Science and Engineering, Uppsala University, SE-751 21 Uppsala, Sweden*

[5]*Division of Materials Physics, Department of Physics and Astronomy, Uppsala University, SE-751 21 Uppsala, Sweden*

[6]*Department of Materials Science and Engineering, Rensselaer Polytechnic Institute, Troy, NY 12180, USA*

[7]*Wallenberg Initiative Materials Science for Sustainability, Department of Chemistry, Uppsala University, 751 21 Uppsala, Sweden*

[*] Corresponding authors: susmita.chowdhury@kemi.uu.se, per.eklund@kemi.uu.se

## S1. Thickness of the samples

The thickness of the samples was obtained from fitting of the X-ray reflectivity (XRR) data. The XRR measurements were performed on a Malvern Panalytical X'Pert Materials Research Diffractometer using CuKα X-ray radiation and 1/32° divergence slit at the incident beam optics side setting the acceleration voltage at 45kV and tube current at 20 mA. A Ni 0.15 mm beam attenuator was also used along the incident beam side setting the activate and deactivate level at 400K and 200K cps to protect the detector from high intense direct beam exposure. With a step size of 0.005°, the scanning range of 2θ was set at 0.2°-3°. During the alignment steps of 2θ→Z→ω→ Z→ω→ψ→ω, the Z height was adjusted for each sample scans due to non-uniformity in exfoliation of the mica layers with the uncontrolled scotch tape procedure.

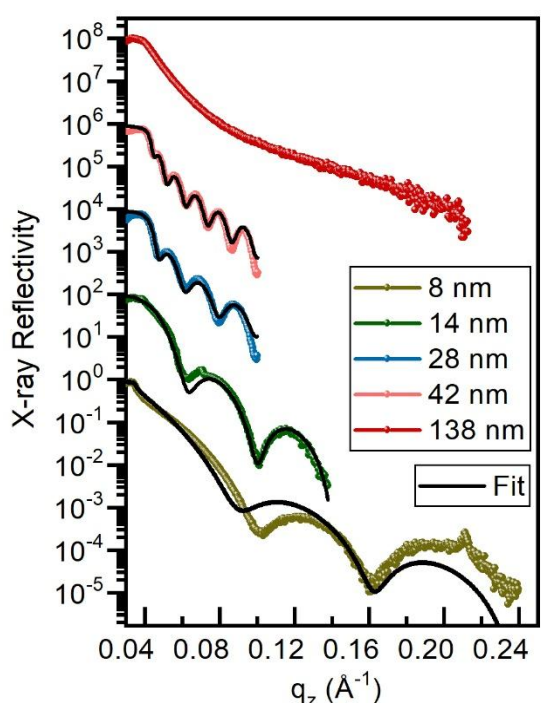

**Fig. S1.** XRR of thickness dependent ScN thin film samples deposited on f-mica substrate.

The XRR fitting was performed using Parratt 32 software based on Parratt's formalism[1]. The Kiessig fringes were observed for all the samples except the thickest sample as shown in Fig. S1. The Kiessig oscillations were not apparent for the thickest sample due to detection limit of the instrument. The thicknesses obtained from the fitted spectra are 8(±1.1) nm, 14(±0.5) nm, 28(±0.2) and 42(±0.2) nm, respectively for the thinner samples and yields a deposition rate of 0.023 nm/s. Hence the thickness of the thickest sample can be estimated at ≈138 nm under the similar deposition conditions except variable deposition time.

## S2. Pole figure of f-mica

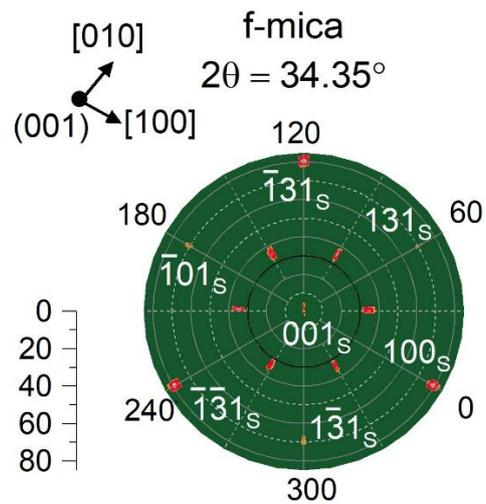

**Fig. S2.** Pole figure in log scale of f-mica substrate.

Fig. S2 shows the pole figure of f-mica measured at a similar Bragg diffraction angle $2\theta = 34.35°$ to that of the ScN film. Four different set of poles were detected from the monocrystalline f-mica substrate at variable χ corresponding to different φ values. They are identified as (i) central pole appears at $\psi = \varphi = 0°$: 001, (ii) six set of poles at $\psi \sim 34.7°$ distributed at $\Delta\varphi \approx 60°$, (iii) three set of poles at $\psi \sim 71°$ and $\Delta\varphi \approx 120°$: 131, $\bar{1}01$ and $1\bar{3}1$ and, (iv) three intense poles at $\psi \sim 79°$ and $\Delta\varphi \approx 120°$: $\bar{1}31$, $\bar{1}\bar{3}1$, 100.

## S3. Compositional analysis of ScN

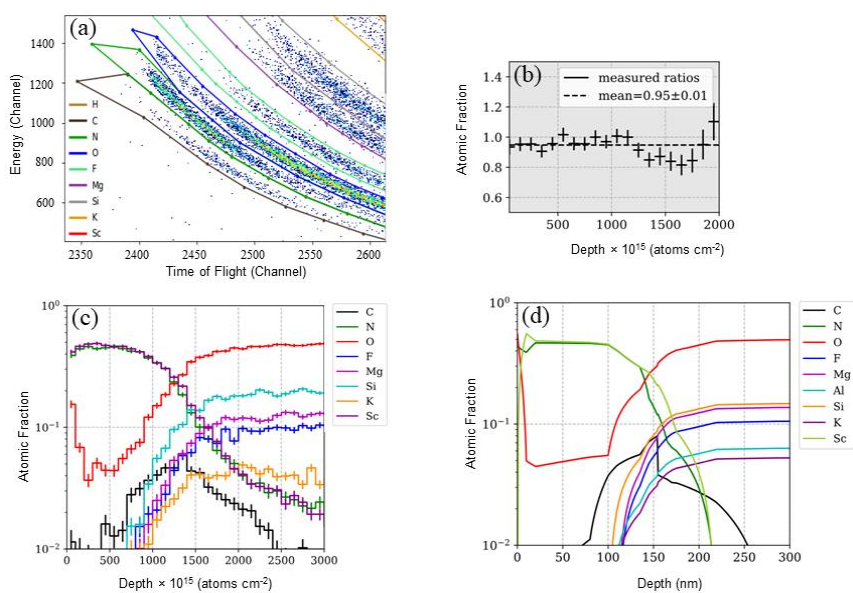

**Fig. S3.** (a) Time-energy histogram, of the as measured ToF-ERDA data, for the ScN film and f-mica substrate; Tracks are labeled according to the element to which they correspond. (b) The N/Sc ratio determined from ToF-ERDA plotted as a function of depth from the surface for the thickest ScN epilayers grown on f-mica. (c) Depth dependent elemental concentrations, as derived by direct data evaluation using Potku, for Sc and N in ScN, and K, Mg, Al, Si, O and F in f-mica along with adventitious O and C. (d) Depth dependent elemental concentrations, as derived by Monte Carlo simulation using MCERD.

The experimental ToF-ERDA data of the thickest ScN film is shown in Fig. S3 a. A mean N/Sc ratio of 0.95±0.01 was obtained for the bulk of the film (Fig. S3 b). The depth dependent elemental concentrations show presence of unintentional O of ~4.5 at. %, ~1.1 at. % C and ~0.2 at. % H within the film (Fig. S3 c). At the film-substrate interface, C contaminations of up to 5 at. % was also detected. The measured and simulated data of Sc, N, O and C shows the good fit of the ToF-ERDA data (Fig. S3 d). The elemental concentrations and their relative uncertainties present in the ScN film are tabulated (Table S1).

Table S1. Elemental concentrations for the bulk of the ScN film, obtained by integrating the ToF-ERDA elemental depth profiles (Fig. S3 a) over a depth range of 200-600×$10^{15}$ atoms cm$^{-2}$. Contributions from F, Mg, Si and K signals were excluded from this integration, under the assumption that they result purely from multiple-scattering effects. This assumption is supported by the MCERD results shown in Fig. S3 d.

| Element | Atomic concentration (at. %) | Absolute Uncertainties (%) |
|---|---|---|
| Sc | 48.2 | 1.01 |
| N | 46.1 | 1.01 |
| O | 4.5 | 0.25 |
| C | 1.1 | 0.12 |
| H | 0.2 | 0.05 |

## S4. Elemental distribution and diffraction pattern of ScN on f-mica

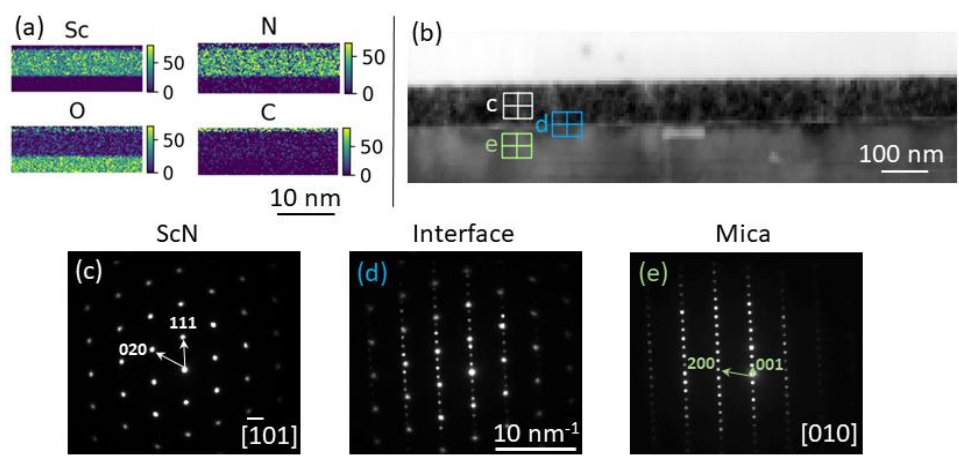

**Fig. S4.** (a) Energy Dispersive X-ray profiles of Sc, N, and unintentional adventitious O and C. (b) From a single 4D-STEM map acquired in nanobeam diffraction mode with indicated ScN film, ScN/mica interface and mica regions, (c-e) integrated diffraction patterns were extracted. The electron diffraction patterns of (c) ScN film, (d) substrate-film interface and (e) f-mica.

Energy Dispersive X-ray Spectroscopy show a homogeneous distribution of N and Sc in the ScN film, as expected (Fig. S4 a). Unintentional presence of adventitious O and C are low in the film. Integrated diffraction patterns were extracted from a single 4D-STEM map (Fig. S4 b) indicating the regions of ScN film, ScN/mica interface and the f-mica, acquired in nanobeam diffraction mode. Spotty electron diffraction patterns from film/substrate interface (Fig. S4 d) consist of superposed patterns from ScN film (Fig. S4 c) and f-mica substrate (Fig. S4 e), indicating crystalline nature of the interface. Further, the ScN thin film is monocrystalline and exempt of any twining boundaries.